\documentstyle[referee,epsf]{mn}
\newif\ifAMStwofonts

\newcommand{\be}{\begin{equation}}
\newcommand{\ee}{\end{equation}}
\newcommand{\bdm}{\begin{displaymath}}
\newcommand{\edm}{\end{displaymath}}
\newcommand{\bea}{\begin{eqnarray}}
\newcommand{\eea}{\end{eqnarray}}

\ifoldfss
  \ifCUPmtlplainloaded \else
    \NewTextAlphabet{textbfit} {cmbxti10} {}
    \NewTextAlphabet{textbfss} {cmssbx10} {}
    \NewMathAlphabet{mathbfit} {cmbxti10} {} 
    \NewMathAlphabet{mathbfss} {cmssbx10} {} 
  \fi
  \ifAMStwofonts
    \ifCUPmtlplainloaded \else
      \NewSymbolFont{upmath} {eurm10}
      \NewSymbolFont{AMSa} {msam10}
      \NewMathSymbol{\upi}     {0}{upmath}{19}
      \NewMathSymbol{\umu}     {0}{upmath}{16}
      \NewMathSymbol{\upartial}{0}{upmath}{40}
      \NewMathSymbol{\leqslant}{3}{AMSa}{36}
      \NewMathSymbol{\geqslant}{3}{AMSa}{3E}

       \let\ge=\geqslant
    \fi
  \fi
\fi 

\ifnfssone
  \newmathalphabet{\mathit}
  \addtoversion{normal}{\mathit}{cmr}{m}{it}
  \addtoversion{bold}{\mathit}{cmr}{bx}{it}
  \newmathalphabet{\mathbfit} 
  \addtoversion{normal}{\mathbfit}{cmr}{bx}{it}
  \addtoversion{bold}{\mathbfit}{cmr}{bx}{it}
  \newmathalphabet{\mathbfss} 
  \addtoversion{normal}{\mathbfss}{cmss}{bx}{n}
  \addtoversion{bold}{\mathbfss}{cmss}{bx}{n}
  \ifAMStwofonts
    \ifCUPmtlplainloaded \else
      \UseAMStwoboldmath
      \makeatletter
      \new@mathgroup\upmath@group
      \define@mathgroup\mv@normal\upmath@group{eur}{m}{n}
      \define@mathgroup\mv@bold\upmath@group{eur}{b}{n}
      \edef\UPM{\hexnumber\upmath@group}
      \new@mathgroup\amsa@group
      \define@mathgroup\mv@normal\amsa@group{msa}{m}{n}
      \define@mathgroup\mv@bold\amsa@group{msa}{m}{n}
      \edef\AMSa{\hexnumber\amsa@group}
      \makeatother
      \mathchardef\upi="0\UPM19
      \mathchardef\umu="0\UPM16
      \mathchardef\upartial="0\UPM40
      \mathchardef\leqslant="3\AMSa36
      \mathchardef\geqslant="3\AMSa3E

       \let\ge=\geqslant
    \fi
  \fi
\fi 

\ifnfsstwo
  \DeclareMathAlphabet{\mathbfit}{OT1}{cmr}{bx}{it}
  \SetMathAlphabet\mathbfit{bold}{OT1}{cmr}{bx}{it}
  \DeclareMathAlphabet{\mathbfss}{OT1}{cmss}{bx}{n}
  \SetMathAlphabet\mathbfss{bold}{OT1}{cmss}{bx}{n}
  \ifAMStwofonts
    \ifCUPmtlplainloaded \else
      \DeclareSymbolFont{UPM}{U}{eur}{m}{n}
      \SetSymbolFont{UPM}{bold}{U}{eur}{b}{n}
      \DeclareSymbolFont{AMSa}{U}{msa}{m}{n}
      \DeclareMathSymbol{\upi}{0}{UPM}{"19}
      \DeclareMathSymbol{\umu}{0}{UPM}{"16}
      \DeclareMathSymbol{\upartial}{0}{UPM}{"40}
      \DeclareMathSymbol{\leqslant}{3}{AMSa}{"36}
      \DeclareMathSymbol{\geqslant}{3}{AMSa}{"3E}

       \let\ge=\geqslant
    \fi
  \fi
\fi 

\ifCUPmtlplainloaded \else
  \ifAMStwofonts \else 
    \def\upi{\pi}
    \def\umu{\mu}
    \def\upartial{\partial}
  \fi
\fi

\title[Reconnection as a cause of
diffuse ionised gas]
{Localized magnetic reconnection as a cause of extraplanar
diffuse ionised gas in the galactic halo}
\author[G.T. Birk et al.]
       {G.T. Birk,$^1$ H. Lesch, $^1$ and T. Neukirch, $^2$\\
        $^1$Institut f\"ur Astronomie und Astrophysik, LMU M\"unchen
      Scheinerstr. 1, 81679 M\"unchen, Germany\\
      $^2$School of Mathematical
and Computational Sciences, University of St. Andrews, St. Andrews,
Fife KY16 9SS, \\
Scotland, United Kingdom}

\date{Accepted
      Received
      }

\pagerange{\pageref{firstpage}--\pageref{lastpage}}
\pubyear{1997}

\begin{document}

\maketitle

\begin{abstract}
Many observations indicate the occurrence of ionised gas in the distant
halos of galaxies (including our own). Since photoionisation by stars
(mainly O stars, young or evolved low-mass stars depending on the kind
of galaxy) does not seem to be exclusively responsible
for the ionisation of the hydrogen filaments that should otherwise
cool fast and recombine quickly, the question arises which extra energy source
can produce the quasistationary ionisation. We show that stationary
localized magnetic reconnection in current filaments may contribute to the
ionisation of the extraplanar halo gas. In these filaments magnetic energy is
dissipated. Consequently, the ionised as well as the neutral component
are heated and re-ionised on a time scale significantly
shorter than the recombination time scale.
The amount of energy required for
efficient re-ionisation can in principle easily be provided by the free
magnetic energy. We present quasi-static models that are characterized by plasma
temperatures and densities that agree well with the observed values for the
diffuse ionised gas component of the interstellar medium. 
Plasma-neutral gas fluid simulations are made to show that the recombination induced
dynamical reconnection process indeed works in a self-regulatory way.
\end{abstract}

\begin{keywords}
Diffuse Ionised Gas,  Galactic Halo,  Magnetic Reconnection
\end{keywords}
\section{Introduction}

During the last decade it became obvious that the existence of a diffuse
ionised gas (DIG) component is of importance for our understanding of the
interstellar medium. This component is widely spread in the Galactic Halo
(cf. Mezger 1978; Reynolds 1990), in the disk halo interface of spiral galaxies
(e.g. Dettmar 1990; Rand et al. 1990) and in irregular, active and
early type galaxies (recent reviews on the detection of DIG
are given by  Walterbos 1991 and Dettmar 1992, 1995). The detection of the DIG
component of the interstellar medium gives rise to the ionisation problem:
What are the energy sources that keep the DIG ionised? In the case of
the extraplanar Galactic DIG, for example,  recombination has to be balanced by
an energy input of about $10^{42} {\rm erg s}^{-1}$ at maximum (Dettmar 1992).
Several suggestions for the solution of this problem can be found in the
literature, among them photoionisation
by O and OB stars (Matthis 1986; Sivan et al. 1986), by white dwarfs
(Sokolowski and Bland-Hawthorn 1991) and by decaying neutrinos
(Sciama and Salucci 1991), shock heating (Sivan et al. 1986)
and microflares (Raymond 1992). In an alternative approach
the existence of DIG is interpreted in terms of the cooling phase
of galactic fountains (Slavin 1993; Shapiro and Benjamin 1993).
However, the proposed processes are seemingly not sufficient to
resolve fully the ionisation problem in the different parameter regimes.
Photoionisation by decaying neutrinos
seems to be inappropriate to explain the measured line ratios (Dettmar and
Schulz 1992). Photoionisation by OB stars also seems to be questionable
for NGC 891 for the same reason (Dettmar and Schulz 1992).
Additionally, special geometric conditions are necessary
to channel the photon flux from the disk O stars to the distant extraplanar
DIG in the Galactic Halo (for a discussion of the problems of different
explanations see Dettmar 1992).

In this contribution we consider a very general process that may help to
solve the ionisation problem: heating and ionisation by magnetic
reconnection. The reconnection process has been proved to be an efficient heating
process in quite different cosmical plasma regimes.
It was shown, e.g., that
this process can explain the X-ray emission of boundary regions of
high-velocity-clouds hitting the Galactic disk (Zimmer et al. 1997a, b),
that it explains the small-scale solar coronal heating in
bright points (Priest et al. 1993; Birk et al. 1997; Dreher et al. 1997), that
it can influence the temperature of accretion disks in active galactic nuclei
(Lesch 1990)
and may result in auroral heating in the Earth's ionosphere (Birk and Otto
1997).

Whereas we feel that this fundamental process may work in all of the mentioned
types of galaxies here we focus on the extraplanar DIG in our Galaxy
in order to obtain reliable quantitative results.
The principal train of thought is as follows. Extraplanar gas
far away from O-stars cools down and recombines effectively.
The recombination results in a lower thermal plasma pressure and in a
higher rate of electron-neutral collisions. This has a two-fold influence
on gas filaments/clouds threaded by an inhomogeneous magnetic field
carrying localized electric currents. For one, the plasma pressure that had
balanced the stresses exerted by the magnetic field is reduced leading to a
contraction of the current carrying region. 
If the total current flowing through this domain is
approximately conserved then this will automatically lead to a higher
current density. Secondly, the increased number of electron-neutral
collisions leads to a reduced electrical conductivity, because
electron-neutral collisions are the dominant contribution to the
conductivity in the halo.
In this situation localized dynamical magnetic dissipation (e.g. Parker 1994)
that can be regarded as a typical reconnection process
is likely to occur. During this process due to the local
violation of ideal Ohm's law  magnetic field lines are reconnected and
magnetic flux is dissipated. The free magnetic energy is mainly converted
into heat via Ohmic dissipation and consequently, the gas is re-heated and can
be re-ionised.

In the next section we present simple stationary solutions
of the ionisation and energy balance equations of this process.
Section 3 is devoted to the application of
stationary reconnection theory to the considered ionisation problem, whereas
in Section 4 we show results of numerical dynamical simulations of the
reconnection process. We sum up and discuss our findings in Section 5.

\section{Quasistatic stationary heating and ionisation by
magnetic reconnection: Plasma density and temperature profiles}

In this section we show that heating and ionisation caused by
magnetic reconnection gives rise to an ionised gas component
in the simplest stationary model available. Though we consider
the basic process to be a dynamical one, the investigation of this simple
stationary model will provide a first rough estimate of the fraction of
ionised gas to be expected.We assume that the velocities
of both the ionised and the neutral component are dynamically of no
importance and thus can be neglected in a first attempt.
We consider a mainly hydrogen
quasineutral partially ionised plasma in ionisation equilibrium
(not necessarily in local thermal equilibrium) in a steady state.
The energetics is assumed to be governed by local Ohmic dissipation during
the reconnection process and radiative losses caused by bremsstrahlung and
recombination radiation.
Accordingly, we have to consider the stationary ionisation
and energy balance equations
\be
\alpha n_p^2 = \iota n_n
\ee
\be
{\bf j}^2 = \sigma (L^{rad}_{brems} + L^{rad}_{recom})
\ee
where $n_p$, $n_n$, ${\bf j}$, $L^{rad}_{brems}$, $L^{rad}_{recom}$
denote the plasma and neutral gas particle density, the current density
and the radiative loss functions (cf. Huba 1994) due to bremsstrahlung
$L^{rad}_{brems}= 1.7 \cdot 10^{-25} n_p^2 T_e^{1/2} {\rm erg} \ {\rm s}^{-1}
{\rm cm}^{-1}$ ($T_e$
is the electron temperature measured in eV) and recombination radiation
$L^{rad}_{recom}= 1.7 \cdot 10^{-25} E_\infty n_p^2 T_e^{-1/2} {\rm erg} \
{\rm s}^{-1} {\rm cm}^{-1}$
($E_\infty$ is the ionisation energy), respectively.
The ionisation frequency $\iota$ is given by
$\iota = 10^{-5} n_p (T_e/E_\infty)^{1/2}exp(-E_\infty/T_e)
(6+T_e/E_\infty) E_\infty^{-3/2} {\rm s}^{-1}$
whereas the recombination rate (dominated by
radiative recombination) is
$\alpha= 5.2 \cdot 10^{-14}(E_\infty/T_e)^{1/2}(0.43+0.5{\rm ln}(E_\infty/T_e)
+0.5(E_\infty/T_e)^{-1/3}){\rm cm}^3 {\rm s}^{-1}$  (cf. Huba 1994).
By $\sigma$ we denote the collisional conductivity due to electron-neutral
collisions (cf. Huba 1994) $\sigma= 2 \cdot 10^{14} n_p e^2 /m_e n_n (k
T_e/m_e)^{1/2} {\rm s}^{-1}$ where $e$, $m_e$ and $k$ are the elementary charge, the
electron mass and the Boltzmann constant.
The local filamentary structure of the magnetic field and the associated
electrical current are assumed to be of the Harris sheet type
(Harris 1962) for simplicity (we assume the magnetic field to vary in
the $y$-direction):
\be
{\bf j} = 7.7 \cdot 10^{-16} \left( \frac{B}{\mu {\rm G}}\right)
\left(\frac{\delta}{{\rm pc}}\right)^{-1} \cosh^{-2}\left( \frac{y}{\delta}\right)
{\bf e}_z \  {\rm statamp \ cm}^{-2}
\ee
where $\delta$ is the length scale of the inhomogeneity of the filamentary
magnetic field, i.e. the half-thickness of the current layer.

In the case of weakly ionised plasmas $n_{t} \approx n_n$, i.e.
filaments of diffuse ionised and relatively dense neutral gas,
we can approximate the plasma density by means of equations (2) and (3):
\be
n_p = \left(2.7 \cdot 10^{-23} \frac{B^2 n_t}{\delta^2
\cosh^4(y/\delta)(1+E_\infty / T_e^{1/4})}
\right)^{1/3} {\rm cm}^{-3}
\ee
and obtain an  equation for the electron temperature
by inserting Eq.(4) in Eq.(1):
\bea
\frac{6.4 \cdot 10^{15} n_t^{2/3} T_e}{E_\infty^{5/2}(6+ T_e E_\infty)}
e^{-E_\infty/T_e}&=&
\left(0.43+0.5{\rm ln}\left(\frac{E_\infty}{T_e}\right)+
0.5\left(\frac{T_e}{E_\infty}\right)^{1/3}\right) \times \nonumber\\
 &\times& \left(\frac{B^2}{\delta \cosh^4(y/\delta)
(1+E_\infty /T_e^{1/4})}\right)^{-1/3}
\eea
that has to be solved numerically. Given a solution for $T_e$
the other parameters, in particular the degree of ionisation $n_p/n_n$,
can be determined.
If we are not dealing with diffuse ionised gas associated with
relatively dense neutral gas filaments but with comparable plasma and
neutral gas densities we instead have to solve for the set of
non-linear algebraic equations:
\be
n_p^3 + 2.7 \cdot 10^{-23} \frac{B^2(n_p - n_t)}{\delta^2 \cosh^4(y/\delta)
(1+E_\infty T_e^{-1/4})} = 0
\ee
\be
\left(0.43+0.5{\rm ln}\left(\frac{E_\infty}{T_e}\right)+
0.5\left(\frac{T_e}{E_\infty}\right)^{1/3}\right) n_p
-2 \cdot 10^8\frac{T_e}{E_\infty^{5/2}}\frac{n_t-n_p}{6+T_e /E_\infty^{1/4}}
e^{-E_\infty/T_e}=0
\ee

The ionisation energy for hydrogen is $E_\infty = 13.6$eV and the lower
estimate for the magnetic
field strength is about $5 \mu$G from radio data (Beck et al. 1996).
The only free input parameters
for our calculations are the total particle density $n_t$ and
the half-thickness of the current sheet $\delta$.
If we assume local thermal equilibrium we can calculate the neutral gas
density by means of the corona model (cf. Huba 1994),
which is applicable for the case of the Galactic Halo, since
$10^{12} \iota n_p < n_p < 10^{16} T_e^{7/2}$ holds:
\be
\frac{n_n}{n_i^2} = \frac{\alpha}{\iota} \approx 80 {\rm cm}^3
\ee

Fig.~1 shows the spatial profiles of the
electron temperature $T_e$, plasma density $n_p$
and neutral gas density $n_n$ for $n_t =1 {\rm cm}^{-3}$ and
$\delta=10^{-11}$pc
(upper row), $n_t = 0.15 {\rm cm}^{-3}$, $n_t = 0.15 {\rm cm}^{-3}$ and 
$\delta=10^{-10}$pc
(middle row) and $n_t =0.5 {\rm cm}^{-3}$ and $\delta=10^{-10}$pc
but a higher magnetic field of $B=30 \mu $G (lower row) that in this case 
may result locally, e.g. from dynamical compression.
In all cases the electron temperature is close to the observed
value of the HII-temperature $T_i \sim 10000$K. The electron and ion
temperatures should not differ significantly since electron-ion collisions
leads to a thermalization on a time scale of $\nu_{ei}^{-1} = 10^{10}{\rm s}
\ll t_{recom}$.
For $n_t=1{\rm cm}^{-3}$ the central ionisation rate is $\sim 60 \% $ which agrees
quite well with observed values in some Galactic DIG regions.
The dependence of the central electron temperature and plasma
density on the length scale $\delta$ is illustrated in Fig.~2.
The lines represent the central plasma density for $n_t =1{\rm cm}^{-3}$
(dashed-dotted), $n_t =0.5{\rm cm}^{-3}$ (solid) and
$n_t =0.15{\rm cm}^{-3}$ (dashed). The central electron temperature is indicated
by triangles (for the case $n_t=1{\rm cm}^{-3}$),
diamonds (for the case $n_t=0.5{\rm cm}^{-3}$) and asterisks
(for the case $n_t=0.15{\rm cm}^{-3}$).

From our idealized model we conclude that magnetic reconnection may account
for the existence of extraplanar ionised gas with observed densities
$n_p \approx 0.01 - 0.5 {\rm cm}^{-3}$ provided that
the magnetic reconnection process takes place in filaments of
the thickness of $\delta =10^{-11} - 10^{-9}$pc.

\section{Stationary heating and ionisation by magnetic reconnection:
Time scales and energetics}

In relatively low resistivity plasmas
the simplest kind of steady magnetic reconnection, Sweet-Parker
reconnection, is usually to slow to account for the observed
phenomena.
The characteristic dynamical dissipation time $\tau_{rec}$ is (cf. Parker
1994):
\be
\tau_{rec} = \frac{\Delta}{v_A} S^{1/2}
\ee
where $S$ is the Lundquist number $S=\Delta v_A/\eta$ and $\Delta$ is the
width of the considered current filaments.
For a half thickness of the current layer of
$\delta = 10^{-9}{\rm pc}$ for our application we obtain
$\tau_{rec} \approx 8 \cdot 10^{19}$s which is much
larger than the recombination time scale
$t_{recomb} = 1/(\alpha n_p) \approx 10^{13}$s (Dettmar 1992), i.e. steady
Sweet-Parker reconnection would not prevent the extraplanar gas from
recombining totally.

However, magnetic reconnection can convert magnetic energy 
into heat on a much faster time scale, if reconnection does not operate
along the entire width of the homogeneous current filament. In our situation
this stronger localization of the dissipative process
may result, e.g. from  slight anisotropies of
recombination along the widths of the current filaments. In this situation
Petschek-like reconnection (Petschek 1964; Parker 1994)
rather than the relatively slow Sweet-Parker
reconnection operates. The velocity of the reconnection process
$u_P$ is given by (Parker 1994):
\be
u_P = \frac{v_A}{{\rm ln} S} = 5.6 \cdot 10^6 u
\ee
and the dissipative region $\Delta^\prime$
is much smaller than $\Delta$ with a lower limit estimated by
\be
\Delta^\prime = \frac{\Delta ({\rm ln} S)^2}{S} = 5 \cdot 10^3 {\rm cm}
\ee
For a dissipative region of the order of the half-thickness
of the current sheet $\delta \sim \Delta$ the reconnection time scale
is $\tau_{rec} = 4 \cdot 10^4$s.
We conclude that a Petschek-like localized reconnection process
can result in heating and thereby ionisation
of the extraplanar gas fast enough to prevent it from recombining totally.
It should be note that the newer generation of reconnection regimes,
i.e. almost uniform reconnection (Priest and Forbes 1986) and nonuniform
reconnection (Priest and Lee 1990) are able to liberate the magnetic energy
even more quickly.

The observational correlation between disk activity (in form of star
formation) and the DIG-brightness in the halo is a clear indicator that
disk kinetic energy is somehow thermalized in the halo gas. The
general scenario suggests that on large scales the interstellar medium is characterized
by energy equipartition between the dynamical
constituents (cosmic rays, turbulence and magnetic fields) (e.g. Ikeuchi 1988).
We have shown above that the dissipation of magnetic
energy via reconnection is able to sustain a local 
recombination-reconnection equilibrium, provided that the 
dissipated magnetic flux is refreshed fast enough, i.e. 
the magnetic dissipation rate
\be
\Lambda_{\rm diss}={B^2\over{8\pi \tau_{\rm rec}}}f_{diss}
\ee
equals the energy input rate caused by disk activity
\be
\Gamma = {\dot E_{\rm disk}\over{V_{\rm diss}}}f_{in}
\ee
where $f_{diss}$ and $f_{in}$ denote the efficiency of magnetic field
dissipation, which is about $f_{diss}=0.1$ for the case of Petschek
reconnection,
and the fraction of conversion of kinetic energy
involved in the disk activity to magnetic fields, respectively.
$\dot E_{\rm disk}$ denotes the kinetic disk luminosity and $V_{\rm diss}$
is the volume in which the following chain of processes takes place:

-- Activity in the disk (e.g. supernovae, stellar winds) gives rise to the
generation of cosmic rays which in course trigger the Parker (buoyancy)
instability, i.e. the disk magnetic field is inflated into the Galactic Halo
(Parker 1994). The rate at which magnetic field is extended and 
magnetic free energy is continuously created is
$10^{41}\, {\rm erg s}^{-1}$.

-- Locally, the magnetic field is compressed until its energy
density is comparable to the kinetic energy density of the plasma.

-- The magnetic field energy is dissipated in filamentary structures
thereby resulting in a re-ionisation of the DIG.

If we assume a reconnection time comparable to the recombination time
($\tau_{\rm rec}\approx 10^{13}\, {\rm s}$), the condition
$\Lambda_{\rm diss}\approx \Gamma$ leads to

\be
V_{\rm diss}\approx {\dot E_{\rm disk} \tau_{\rm rec}8\pi \over B^2}
\frac{f_{in}}{f_{diss}}\approx
10^{67} {\rm cm}^3 \frac{f_{in}}{f_{diss}}
\left[{\dot E_{\rm disk}\over{10^{41}\, {\rm erg s}^{-1}}}\right]
\left[{\tau_{\rm rec}\over{10^{13}\, {\rm s}}}\right]
\left[{B\over {5\mu {\rm G}}}\right]^{-2}
\ee

which corresponds to a sphere with a radius of about 7 kpc for
$f_{in}/f_{diss}=10$.
In this volume the magnetic energy should be dissipated in a huge number of
neighbored individual (observationally non-resolved) reconnection processes as 
described above.
We note that, depending on the actual type of reconnection, the individual
heating processes are not restricted to the entire region of the current 
filaments but efficient heat transport, e.g. due to outflow and
shocks, on larger spatial scales is to be expected (e.g. Biskamp 1993).
Additionally, the individual reconnection events excite Alfv\'en waves 
which can dissipate the energy over lager spatial scales (Champeaux et al.
1997).
The scenario we have in mind  resembles very much the concept of coronal heating by 
nano flares (Parker 1994; cf. also discussion
by Raymond 1992).

\section{Dynamical heating and ionisation by magnetic reconnection:
Numerical Simulations}

Finally, we can drop the explicit assumption of a stationary configuration and
model the ionisation of the DIG by reconnection
with the help of numerical dynamical simulations. 
These simulations are meant to show that the proposed reconnection scenario
indeed can be regarded as a dynamical self-regulating process.
The balance equations (note that we are not dealing with standard MHD
equations; our analysis rather makes use of a multi-fluid description) that are
integrated by means of an explicit difference scheme (for details of the
numerical procedure see Birk and Otto 1997) read:
\be
{\partial \rho \over \partial t} = - \nabla \cdot ( {\bf v} \rho)
+ \iota \rho_n - \alpha \rho^2
\ee

\be
{\partial \rho_n \over \partial t} = - \nabla \cdot ( {\bf v}_n \rho_n)
- \iota \rho_n + \alpha \rho^2
\ee

\be
{\partial \over \partial t} ( \rho {\bf v}) = - \nabla \cdot ( \rho {\bf v  v} )
 - \nabla p +  (\nabla \times {\bf B}) \times
{\bf B}  -  \rho \nu_{pn} ({\bf v} - {\bf v}_n)  + \iota \rho_n {\bf v}_n
- \alpha \rho^2 {\bf v}
\ee

\be
{\partial \over \partial t} ( \rho_n {\bf v}_n) = - \nabla \cdot
( \rho_n {\bf v_n v_n} )
 - \nabla p_n  
- \rho_n \nu_{np} ({\bf v}_n - {\bf v}) - \iota \rho_n {\bf v}_n
+ \alpha \rho^2 {\bf v}
\ee

\be
{ \partial {\bf B} \over \partial t } = \nabla \times
({\bf v} \times {\bf B})
- \nabla \times ( \eta \nabla \times {\bf B} )
\ee

\bea
{\partial p \over \partial t} = &-& {\bf v} \cdot \nabla p
-\gamma p \nabla \cdot {\bf v}
+ (\gamma -1) \biggl[ \eta (\nabla \times
{\bf B})^2  -  \nu_{pn}
\left( p - {\rho \over \rho_n} p_n \right)
+\iota p_n - \alpha \rho p \nonumber \\
&+&  (\rho \nu_{pn} +{1 \over 2}(\iota \rho_n + \alpha \rho^2))
({\bf v} - {\bf v}_n)^2 \biggr]
\eea

\bea
{\partial p_n \over \partial t} =
&-& {\bf v}_n \cdot \nabla p_n  - \gamma_n p_n  \nabla \cdot {\bf v_n}\nonumber \\
&-&  (\gamma_n - 1) \biggl[
\nu_{np} \left( p_n - {\rho_n \over \rho} p \right)
+\iota p_n - \alpha \rho p
+  (\rho_n  \nu_{np} +{1 \over 2}(\iota \rho_n + \alpha \rho^2))
({\bf v}_n - {\bf v})^2 \biggr]
\eea
In this set of equations $\rho$, ${\bf v}$,
$p$, $\nu_{pn}$, $\nu_{np}$, $\gamma$ and $\gamma_n$
denote the mass density, the bulk velocity, the thermal pressure,
the elastic plasma-neutral gas collision frequencies ($\nu_{pn} \rho
= \nu_{np} \rho_n$) and the ratios of the specific heats.
The initial configuration is characterized by an ideal
plasma and an isothermal homogeneous neutral gas in ionisation equilibrium.
The Harris-type current sheet (${\bf B} \sim tanh(y) {\bf e}_x$)
is balanced by thermal plasma pressure ($p\sim sech^2(y)$).

The dimensions of our numerical box are given by $x \in [-10,10]$ and
$y \in [-10,10]$ in normalized units
(half-width of the current layer).
The simulations are carried out with $77$ grid points in both
directions.
The numerical grid is chosen non-uniform in the $x$-direction
with a maximum solution of $0.05$.

The initial equilibrium is perturbed by a localized enhancement of the
recombination coefficient $\alpha_{pert}\sim sech^2(x)sech^2(y)$.
This perturbation results in a decrease of the thermal plasma pressure
and thereby in an inward plasma transport (Fig.~3, upper panel). 
Magnetic field lines
are convected inward, since the frozen-flux condition holds outside the
central current filament (Fig.~3, lower panel). A dissipative region
forms inside the filament due to the enhanced recombination.
The formation of this non-ideal region is modeled by a
localized magnetic diffusivity switched on during the temporal evolution.
After $100$ Alfv\'enic times ($\tau_A=\sqrt{4 \pi \rho }\delta/B$)
well developed X-type reconnection results in localized
dynamical magnetic dissipation and divergent plasma flow
caused by Lorentz forces (Fig.~4). The flow and magnetic patterns have
the appearance of flux pile-up reconnection (Priest and Forbes 1986).

During the reconnection process magnetic energy is partly converted into
heat via Ohmic dissipation (cf. Fig.~5). Thermalization due to plasma-neutral 
collisions leads to an increase of the central neutral gas temperature 
(see equation (21)). Eventually, the neutral gas will be re-ionised by the 
local magnetic dissipation mechanism. 

Our numerical studies show that the recombination induced reconnection indeed
can work as a dynamical self-regulating process that, in principle, can keep
a significant part of the interstellar medium in an ionised state.

\section {Summary and Discussion}

We have suggested a magnetic reconnection model for the solution of the
ionisation problem. Similar to Raymond (1992) we assume that the dissipation
of free magnetic energy, stored in thin current filaments, plays a crucial role in
providing energy input necessary for keeping the DIG ionised.
At first sight the existence of such dynamic filaments seems to be a 
little speculative, since due to restrictions of resolution
there are presently no direct observational indications for them.
However, there is strong evidence for a filamentary structure of the ISM from the 
Westerborg survey (Hartmann and Burton 1996) with the filamentary
scale lengths appearing to be restricted by the highest resolution, only.
We assume that even higher resolution would show finer structures.
Additionally, very fine non-thermally emitting radio filaments, and even
substructures of these filaments, have been
observed near the Galactic center (Morris 1996 and ref. therein).
A very nice example for the omnipresence of filamentary gas structure
is the Orion nebula. Subtraction of the continuous spectrum reveals M43
to be a beautiful tangle consisting of astonishingly narrow 
string like ionized filaments (Yusef-Zadeh 1990).
What is more, magnetized plasmas in general seem to have a strong tendency to
show a filamentary structure (cf. also discussion by Alfv\'en 1981)
as has become obvious from the investigations of plasma systems that can 
be observed more closely, e.g. in the auroral ionosphere (e.g. Borovsky and
Suszcynsky 1993; L\"uhr et al. 1994) and the solar corona (cf. Parker 1994).
As a matter of fact, the more the observational resolution improves the more
evident becomes the filamentation aspect.

Our analysis was carried out in the framework of fluid theory.
This should be justified for the calculations given in Sect.~3, in particular,
where the assumption of ionisation equilibrium implies a magnetohydrodynamic
approach. This approach prove appropriate, one the hand, because the
smallest length scale involved $\delta=10^{-9}$pc is much larger than the
ion gyro radius $r_i=v_i/\Omega_i=6\cdot10^{-12}$pc ($v_i=\sqrt{kT_i/m_i}$
is the ion thermal velocity and $\Omega_i=eB/m_ic$ is the ion gyro frequency)
even for the lower estimate of the magnetic field strength $B=5\mu$G.
On the other hand, the collective behavior is guaranteed even in the almost
collisionless regime outside the central current sheets, since $v_A\ge v_i$,
where $v_A = B/\sqrt{4\pi\rho_i}$ is the Alfv\'en velocity, i.e. in this case
typical wave phenomena rather than collisions (as in ordinary hydrodynamics)
guarantee the correlation over macroscopic length scales.
The latter condition is fulfilled somewhat marginally $v_A \approx v_i
\approx 10^6 {\rm cm/s}$ for the lower
estimate of the magnetic field strength and the upper limit of the considered
mass density ($\rho=1{\rm cm}^{-3}$), i.e. the worst case in the present context.
However, even for the case $v_A<v_i$  a fluid
description can still work in a satisfactory way, as can be seen, e.g. 
in the context 
of magnetospheric physics (e.g. Schindler and Birn 1986; 
Baumjohann and Treumann 1997).
Additionally, it should be noted that the magnetic reconnection process 
itself also works in a completely collisionless regime described by
kinetic theory (e.g. Biskamp et al. 1995).

We presented stationary quasistatic solutions that describe the
dissipation process. From these solutions we conclude that
magnetic dissipation can keep the interstellar medium partially ionised with
temperatures of about $1$eV.
We addressed the question whether the conversion of magnetic energy
is fast enough to keep the DIG ionised. We applied stationary reconnection
to show that a sufficient amount of stored energy can be
dissipated on a time scale shorter than the recombination time
for reasonable physical parameters.
Numerical studies support and illustrate our model.
Obviously, our approach cannot rule out other re-heating mechanisms,
such as photoionisation and wave as well as shock heating, 
but it shows that magnetic reconnection should be considered as a
quite natural alternative and additional process that, in principle, can
cause re-ionisation in all kinds of magnetized galactic DIG structures.

Here we assumed that the velocities of the plasma and neutral gas are
both dynamically unimportant, so that the recombination-reconnection
circle is only intrinsically driven. There is also the possibility that 
reconnection may be externally driven by plasma motions.

We note that this possibility and the important role
of the magnetic field for the DIG is clearly indicated by the
observational well established spatial correlation between the
$H\alpha$-distribution and the intensity and polarization distribution
of the nonthermal radio continuum (Dettmar 1992). Since the radio
continuum is due to synchrotron radiation the DIG appears in a magnetized
environment, which is driven by the disk activity via star formation,
supernova winds etc. These processes transport magnetic field and plasma
into the halo and the deposited kinetic energy may then be dissipated in the
halo via externally driven magnetic reconnection and subsequent recombination.

Finally we want to mention that the reconnection-recombination
circle may be applicable to several other astrophysical situations
such as cooling flows and the internal dynamics of molecular clouds, for example.

\section{Acknowledgments} This work was supported by the Deutsche
Forschungsgemeinschaft through the grant ME 745/18-1 and by the United
Kingdom Particle Physics and Astronomy Research Council.

{}

\newpage

\begin{figure}
\caption{Spatial profiles of the electron temperature (measured in eV), 
the plasma and the
neutral gas density for total densities of $n_t =1{\rm cm}^{-3}$ (upper row),
$n_t=0.15{\rm cm}^{-3}$ (middle row) and $n_t=0.5{\rm cm}^{-3}$ 
but $B=30\mu {\rm G}$ (lower row).}
\label{figure1}
\end{figure}

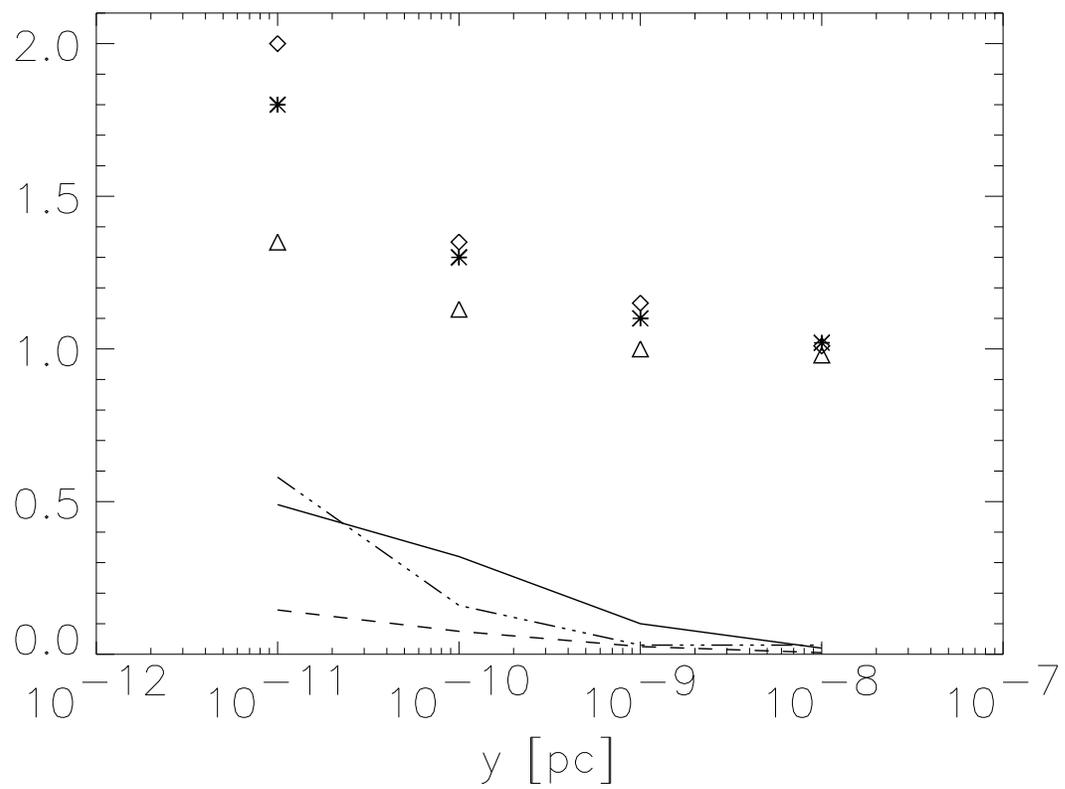
\begin{figure}
\caption{The central plasma density for $n_t =1{\rm cm}^{-3}$ (dashed-dotted line),
$n_t =0.5{\rm cm}^{-3}$ (solid line)
and $n_t =0.15{\rm cm}^{-3}$ (dashed line) and the central electron temperatures
(triangles for $n_t =1{\rm cm}^{-3}$,
diamonds for $n_t=0.5{\rm cm}^{-3}$ and asterisks for $n_t=0.15{\rm cm}^{-3}$)
as functions of the half-width of the current filament}
\label{figure2}
\end{figure}

\begin{figure}
\caption{The plasma velocity (upper graph) and magnetic (lower graph) field
that evolve after $t=10\tau_A$. Plasma flows inward due to the reduced thermal
plasma pressure caused by localized enhanced recombination and magnetic flux is
convected towards the dissipative region.}
\label{figure3}
\end{figure}
\begin{figure}
\caption{The velocity (upper graph) and magnetic (lower graph) field
that evolve after $t=100\tau_A$. Plasma is accelerated from the X-like
reconnection site where the magnetic field is dynamically dissipated.}
\label{figure4}
\end{figure}

\begin{figure}
\caption{The Ohmic heating rate (upper graph), the plasma temperature (middle graph)
and neutral gas temperature (lower graph) in normalized units (the initial
plasma and neutral gas temperatures are chosen $1$ in normalized units)
after $t=100\tau_A$. Plasma is heated via Ohmic dissipation. Frictional
heating results in heat transfer to the neutral gas which in the course will be
ionised.}
\label{figure5}
\end{figure}

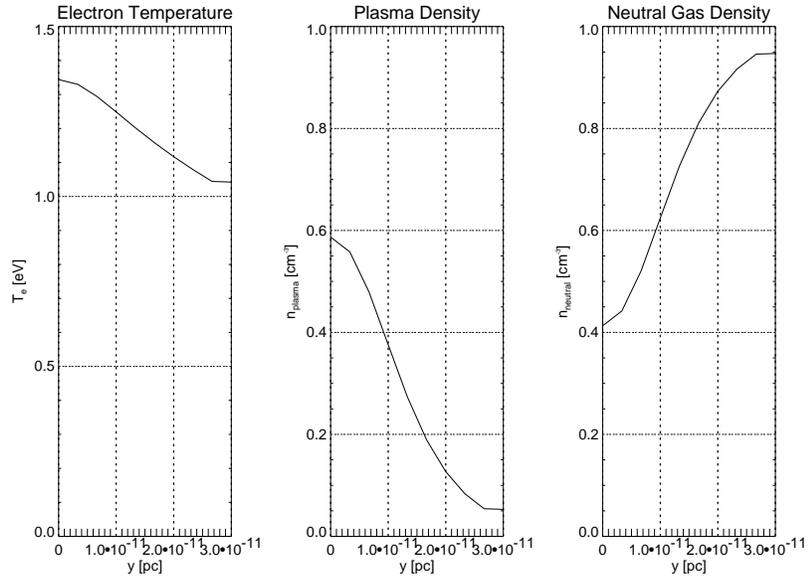
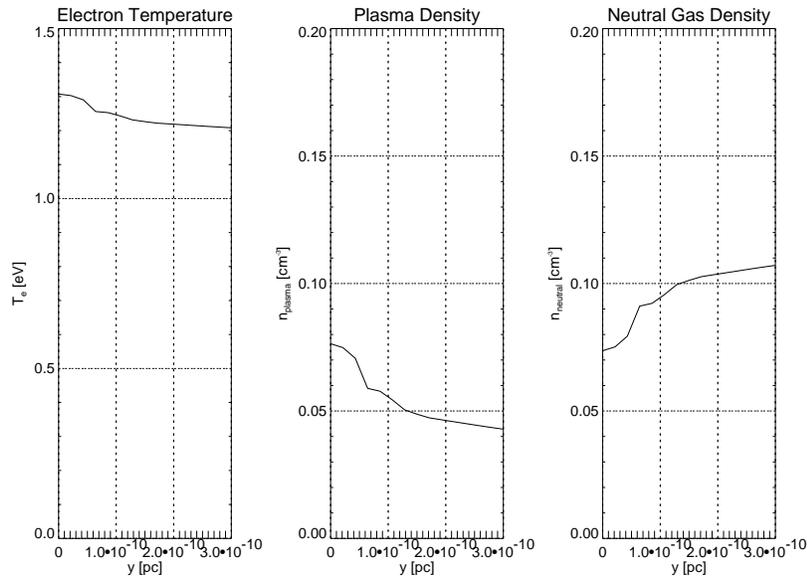
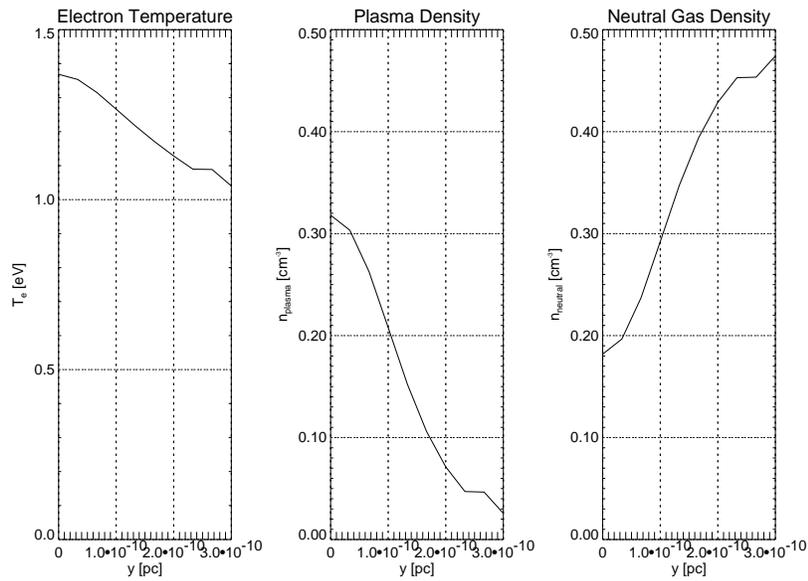
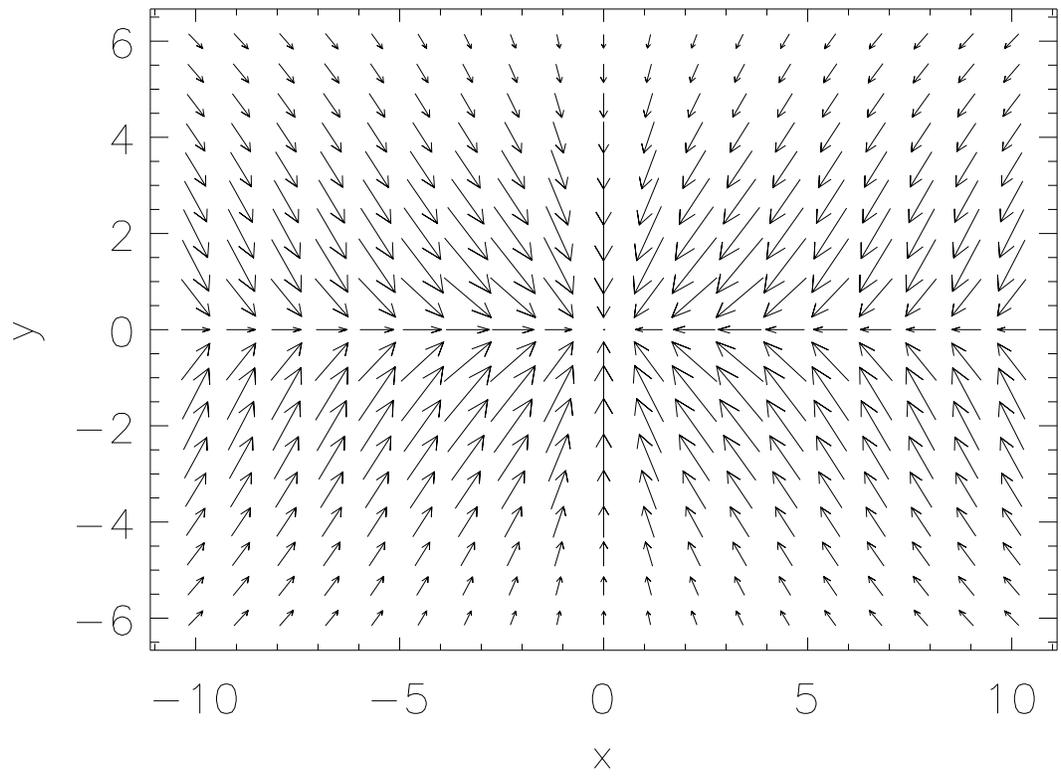
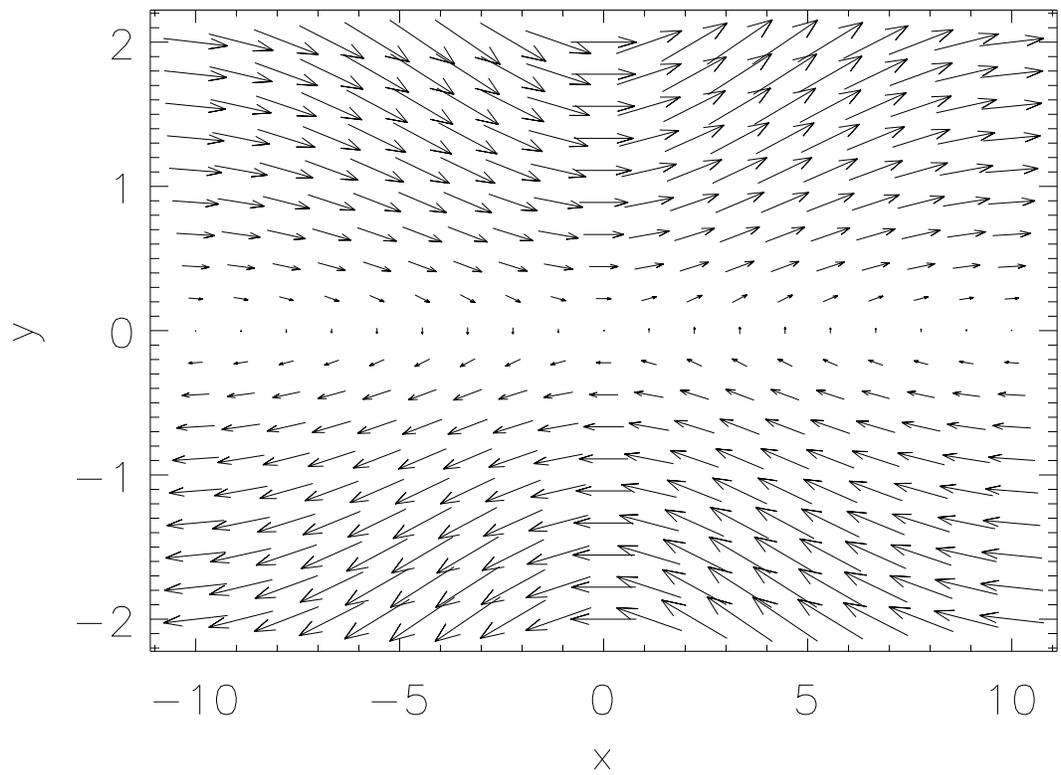
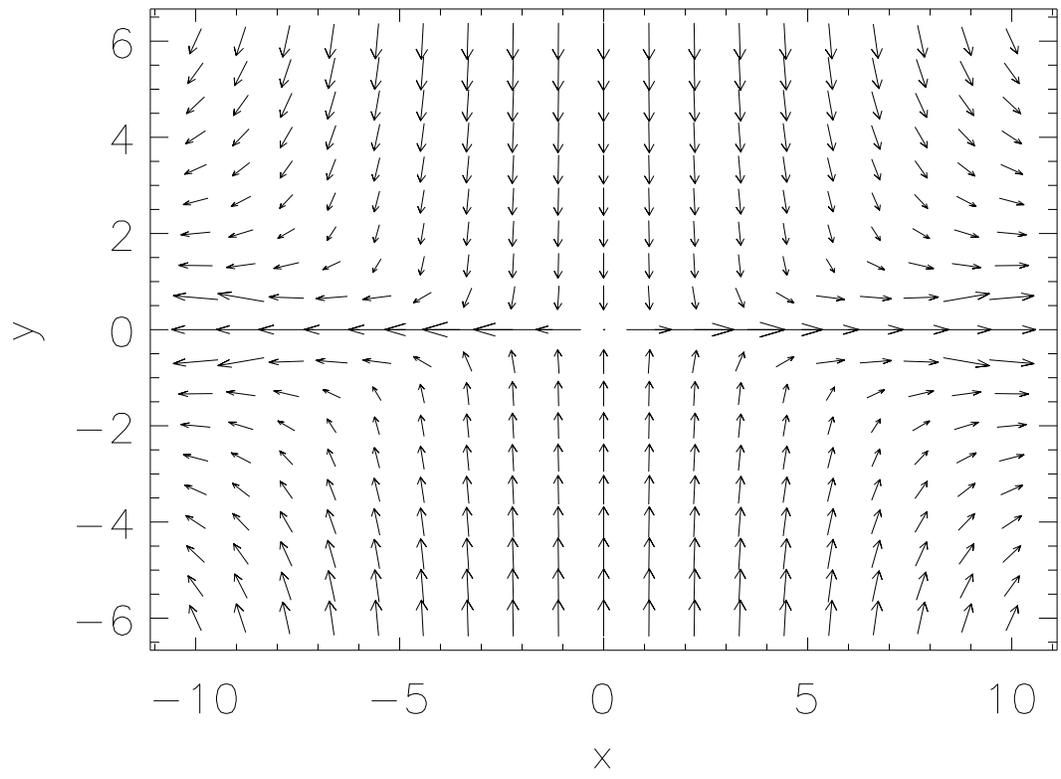
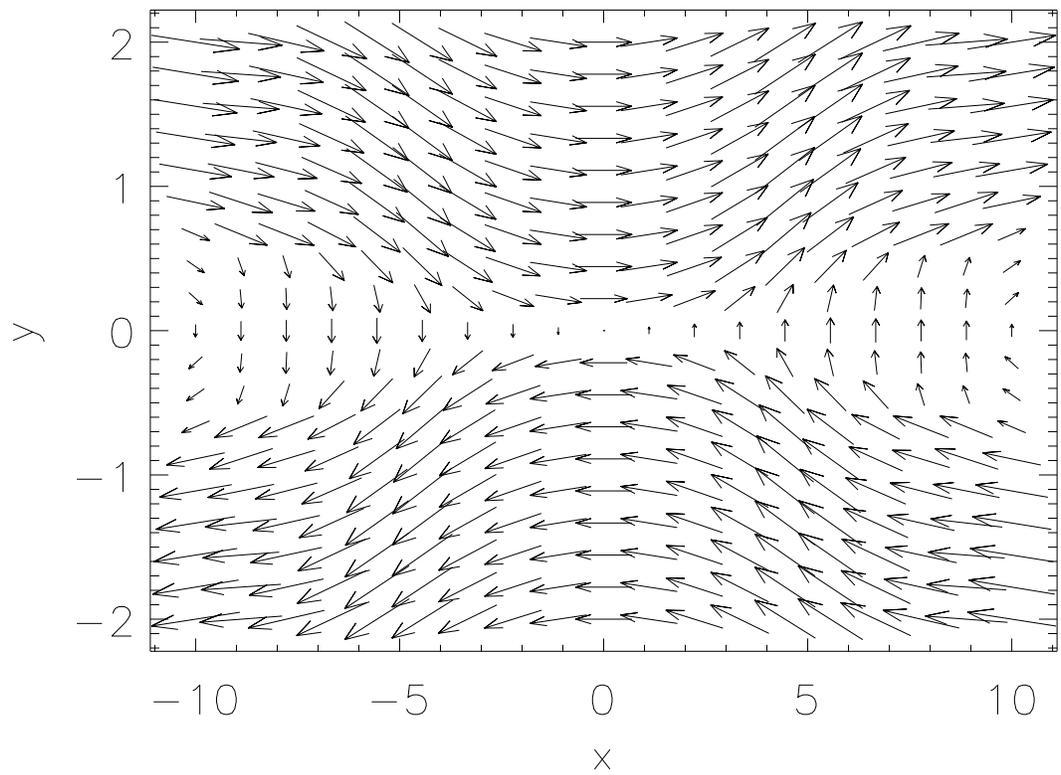
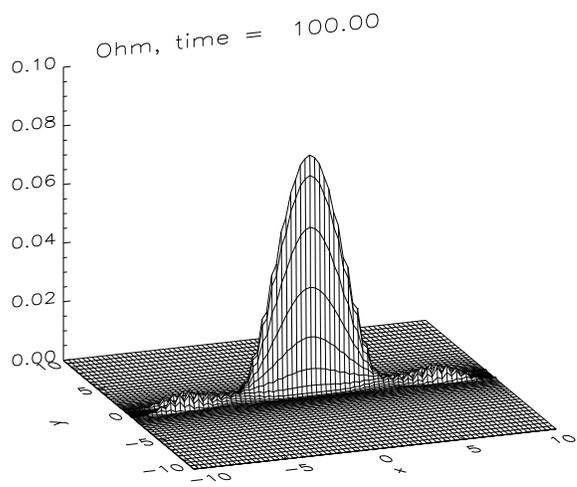
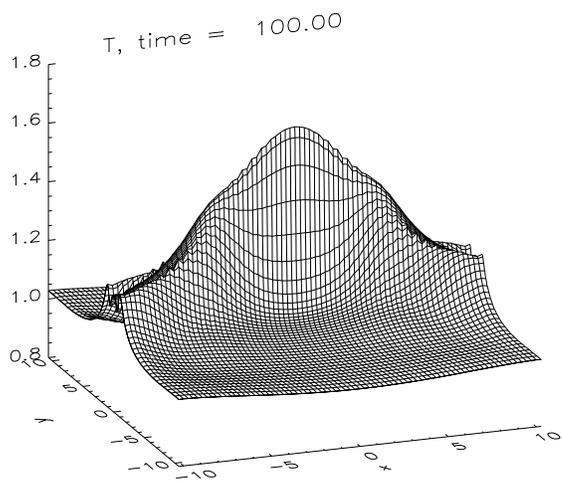
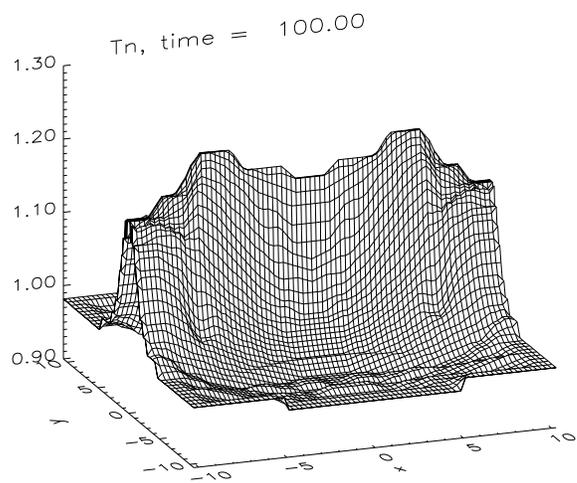
\end{document}